\documentclass[final,3p,times,twocolumn]{elsarticle}

\usepackage{amssymb}
\usepackage{xcolor}
\usepackage{amsmath}

\journal{Optics Communications}

\begin{document}

\begin{frontmatter}



\title{Output beam shaping of a multimode fiber amplifier}


\author[inst1,inst5]{Stefan~Rothe}
\author[inst2,inst5]{Kabish~Wisal}
\author[inst1,inst3]{Chun-Wei~Chen}
\author[inst1]{Mert~Ercan}
\author[inst4]{Alexander~Jesacher}
\author[inst2]{A.~Douglas~Stone}
\author[inst1,inst6]{Hui~Cao}

\affiliation[inst1]{organization={Department of Applied Physics},
            addressline={Yale University}, 
            city={New Haven},
            state={CT},
            postcode={06520}, 
            country={USA}}

\affiliation[inst2]{organization={Department of Physics},
   addressline={Yale University}, 
            city={New Haven},
            state={CT},
            postcode={06520}, 
            country={USA}}

\affiliation[inst3]{organization={Edward L.~Ginzton Laboratory},
            addressline={Stanford University}, 
            city={Stanford},
            state={CA},
            postcode={94305},
            country={USA}}
            
\affiliation[inst4]{organization={
Institute
of Biomedical Physics},
            addressline={Medical University of Innsbruck}, 
            city={Innsbruck},
            postcode={6020},
            country={Austria}}
            
\affiliation[inst5]{organization={These authors contributed equally}}

\affiliation[inst6]{organization={Corresponding author: hui.cao@yale.edu}}

\begin{abstract}
	Multimode fibers provide a promising platform for realizing high-power laser amplifiers with suppressed nonlinearities and instabilities. The potential degradation of optical beam quality has been a major concern for highly multimode fiber amplifiers. We show numerically that the beam propagation factor $M^2$ of a single-frequency multimode fiber amplifier can be reduced to nearly unity by shaping the input or output beam profile with spatial phase-masks. Our method works for narrowband multimode fiber amplifiers with strong gain saturation, pump depletion, random mode coupling and polarization mixing.  The numerical results validate our approach of utilizing highly multimode excitation to mitigate nonlinear effects in high-power fiber amplifiers and performing input wavefront shaping to control output beam profile and polarization state.
 
\end{abstract}




\end{frontmatter}


\section{Introduction}


High-power fiber laser amplifiers~\cite{jeong200443, jeong2007power, richardson2010high, nilsson2011high, jauregui2013high, zervas2014high, dong2016fiber} have a wide range of applications from material processing~\cite{Welding2018} to directed energy~\cite{Defense2019}. 
	Further power scaling of fiber amplifiers \cite{dawson2008analysis} is currently limited by nonlinear optical effects such as stimulated Brillouin scattering~(SBS)~\cite{boyd2020nonlinear, kobyakov2010stimulated, wolff2021brillouin} and transverse mode instability~(TMI)~\cite{Cesar2020transverse, eidam2011experimental, smith2011mode, Ward2012origin}. Various techniques have been developed to mitigate these effects, but most of them are only applied to single-mode or few-mode fibers to ensure output beam quality~\cite{eidam2011preferential,robin2014modal,boyd2020nonlinear,hawkins2021kilowatt,shi2022700}.  

Recently, we proposed and showed theoretically that highly multimode fiber~(MMF) amplifiers can have much higher power thresholds of SBS and TMI than their single-mode counterparts~\cite{chen2023suppressing, wisal2024theorySBS, wisal2024theoryTMI, wisal2024optimal},
when many fiber modes are coherently excited. We also demonstrated experimentally an order of magnitude increase of the SBS threshold in passive step-index MMF over the standard SMF~\cite{chen2023mitigating}. The SBS suppression results not only from reduction of light intensity in a large-core fiber but also from broadening of the Brillouin spectrum by multimode excitation~\cite{wisal2024theorySBS}. Our scheme does not require specialized fibers, thus simplifying implementation and making a MMF amplifier a promising alternative for exploration.
However, output beam quality has been the main concern for employing MMFs, as without any further beam shaping the spatial intensity profile is speckled and difficult to focus or collimate with a lens~\cite{zervas2014high}. A key feature of our approach is adding a beam-shaping apparatus to convert the speckled beam generated naturally by propagation within the fiber to a smooth beam of high quality. 

Laser beam quality is usually characterized by the beam propagation factor $M^2$~\cite{siegman1998maybe, ross2006appropriate}. Its value is given by the normalized product of the second moments of time-averaged intensities in the beam waist and the far field~\cite{siegman1990new, siegman1993defining}. It determines the focusability and divergence of an optical beam. A large $M^2$ limits not only how tightly the output beam of a MMF can be focused by a lens, but also how well it can be collimated to the far field. A diffraction-limited lowest-order Gaussian beam has the minimum value of $M^2 =1$. In an optical fiber, the beam propagation factor is almost unity for the fundamental mode~\cite{belanger1993beam,yoda2006beam}, but becomes significantly larger for higher-order modes~\cite{saghafi1998beam, yoda2006beam, jeong2009multi}. Interestingly, for a coherent superposition of multiple modes, $M^2$ varies substantially with the relative phase of these modes~\cite{yoda2006beam, wielandy2007implications}, which can be exploited to achieve small $M^2$  even when higher-order modes are excited, as we show below. 

Although the $M^2$ of a beam is invariant with beam propagation in free-space and through unaberrated
optical elements, the $M^2$ value can be changed by optical aberration or spatial phase modulation. Such common imperfections in the optical system
increase $M^2$, which is undesired. However, for a coherent beam, time-reversal symmetry
implies that spatial phase modulation can, in principle, shape a beam with large $M^2$ “back” to one with
small $M^2$. In holography the power of spatial phase modulation is well-known to imply that any far-field
intensity pattern for a coherent light can be constructed from a near-field phase hologram. The resulting change of the far-field second-order moment can modify $M^2$ and potentially reduce it dramatically.

In the case of MMFs, the reversibility of an increase in $M^2$ can be seen as follows. Consider a linear MMF with random mode coupling and negligible loss, into which is input a monochromatic Gaussian beam with $M^2 = 1$; the resulting output will be a speckle pattern with $M^2 \gg 1$. However, phase conjugating the output field and sending it back to the same MMF will recover the original Gaussian beam at the fiber input. Thus it clearly is possible to convert a speckled pattern to a Gaussian beam and reduce $M^2$ to unity. 

For practical applications, however, optical phase conjugation and back propagation through the same MMF may be inconvenient or even impossible; instead spatial light modulation is the most convenient way to shape the output beam of a MMF. It is not clear whether, and under what conditions, this wavefront shaping can convert a speckled output beam to a diffraction-limited Gaussian beam. Further questions are (i) whether such conversion is always possible for any coherent superposition of fiber modes, regardless of the number of modes and strength of random mode coupling and polarization mixing in the MMF; (ii) whether the dramatic reduction of $M^2$ will necessitate significant power loss. 
	
Regarding power handling of MMF amplifiers, spatial phase modulation of a low-power seed is clearly more convenient than modulating the high-power output beam. It has been shown that input wavefront shaping can control the output spatial profile, temporal pulse shape, polarization state of MMFs (see Ref.~\cite{cao2023controlling} and references therein). While most of the relevant works were conducted with passive fibers, output beam focusing, steering, and shaping were demonstrated in active MMFs with weak gain saturation~\cite{florentin2017shaping, florentin2018space, florentin2019shaping}. However, it has not been shown that input wavefront shaping can produce a diffraction-limited output beam from a nonlinear MMF amplifier with strong gain saturation.  
	
In this paper, we first show numerically that the speckled output beam of a MMF with $M^2 \gg 1$ can be converted to a diffraction-limited Gaussian beam with $M^2 \approx 1$
by diffracting the output beam consecutively from spatially-separated phase-masks. This process, in principle, does not cause any power loss. Such a method works for an arbitrary coherent superposition of modes in a MMF, with any degree of uncontrolled mode coupling and polarization mixing in the fiber. We further show that even in a high-power MMF amplifier with strong gain saturation and random mode coupling, a speckle-free beam with $M^2 \approx 1$ can be generated by shaping a coherent \emph{input} beam with three phase-masks for each polarization. 
	
Our approach is valid for coherent light with spectral bandwidth narrower than the spectral correlation width of the MMF when all modes are excited to mitigate nonlinear effects. The width characterizes how fast the output field distribution decorrelates with frequency detuning for a fixed input wavefront to the fiber. It limits the allowable bandwidth of the seed to a MMF amplifier. For passive MMFs, the spectral correlation width decreases with increasing fiber length and mode/polarization dispersion. We show numerically that the spectral correlation width is not altered significantly by gain saturation and pump depletion in a nonlinear MMF amplifier. All of our studies are performed via extensive numerical simulations of MMF amplifiers with realistic parameters, taking into account gain saturation, pump depletion, mode-dependent gain, random mode coupling, and polarization mixing. Our results confirm that it is possible to create $M^2 \approx 1$ output beams from MMF amplifiers, removing the beam quality as a major concern, and paving the way for development of highly MMF amplifiers for further power scaling beyond what is achievable with SMF amplifiers.

\section{Output wavefront shaping}
\label{sec:sec2}

	\begin{figure*}[!h]
		\centering
  \includegraphics[width=0.9\textwidth]{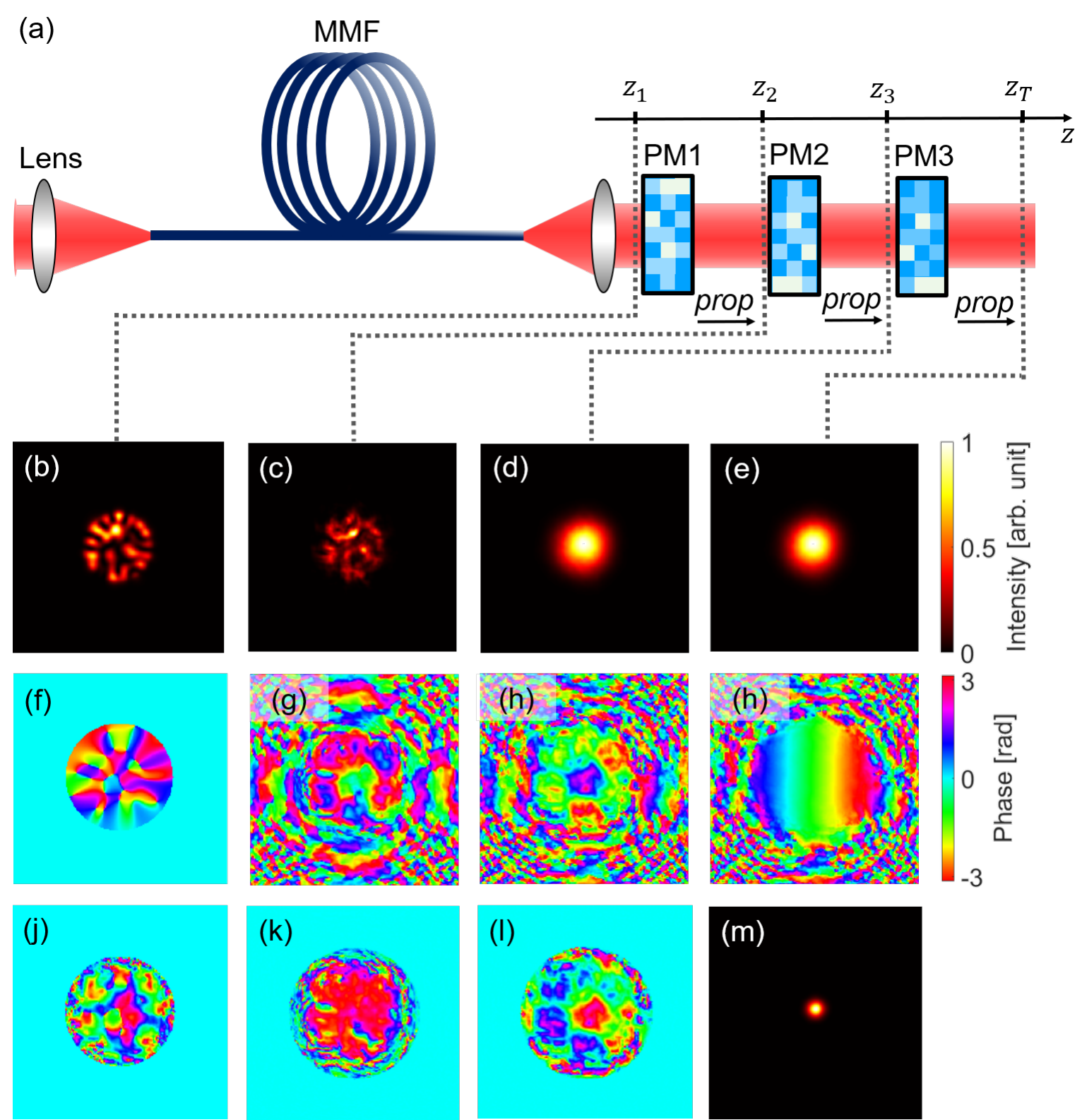}\\
		\caption{Wavefront shaping of a MMF output beam. (a) A schematic of wavefront shaping setup we numerical simulate. The MMF of core diameter 42~\textmu m and NA 0.1 supports 40 transverse modes per polarization at wavelength $\lambda$ = 1064~nm. All modes are equally excited. The output beam displays speckled intensity pattern (b) and random phase distribution (f).  The beam propagation factor is $M^2=6.18$.  After passing through three phase-masks (PM1, PM2, PM3) at $z= z_1, z_2, z_3$, the speckled beam transforms to a Gaussian beam with $M^2=1.17$ in (e). Light intensity patterns (b-e) and phase distributions (f-i) before three phase-masks and at the target plane $z_T$. (j)-(l): spatial phase modulations on three masks obtained by the gradient descent optimization. (m): far-field intensity distribution calculated by Fourier transform of the field distribution at the target plane $z_T$. }
		\label{fig:fig1}
	\end{figure*}

	We start with a passive MMF with random mode coupling induced by fiber imperfections and/or external perturbations. Such coupling is linear, and assumed to be static and deterministic. Figure~1a is a schematic illustrating the conversion process at the fiber output end in order to undo the random mode coupling and generate a smooth beam with small $M^2$. When a spatially and temporally coherent beam is launched to the MMF, the output is a speckle pattern formed by multimode interference. We simulate a MMF of core diameter 42 \textmu m and NA 0.1. At the wavelength $\lambda$ = 1064 nm, the number of guided modes is $N=40$ per polarization~\cite{snyder1978modes}. 
    For now we ignore polarization mixing in the fiber and will include it later in section 4. With the input field linearly polarized along the horizontal 
    axis, the output remains linearly polarized in the same direction.  To mitigate nonlinear effects such as SBS in the fiber \cite{chen2023suppressing, wisal2024theorySBS, wisal2024theoryTMI, wisal2024optimal}, all guided modes are excited equally. The output field is a linear superposition of all 40 modes with equal amplitude and random phase. The output beam propagation factor $M^2 = 6.18$.   

    We then design a series of phase-masks at the fiber output end. The speckled beam from the fiber passes through them with free-space propagation in between. Such concatenated phase modulations will transform any field profile to the desired one. This is similar to the multi-plane light conversion~(MPLC) technique that has been employed in recent years for various applications such as high-dimension mode multiplexing and sorting~\cite{labroille2014efficient,labroille2017characterization, fontaine2017design, fontaine2019laguerre,fontaine2021hermite,rademacher2021peta, kupianskyi2023high, wen2021scalable, butaite2022build, pohle2023intelligent, korichi2023high}, coherent beam combining~\cite{billaud2019optimal}, processing of entangled photons in high dimensions~\cite{brandt2020high,lib2022processing}, rotation sensing~\cite{eriksson2023sensing}, and untangling light propagation through MMFs~\cite{kupianskyi2024all}, as well as diffractive neural networks based on consecutive phase modulators for computational imaging applications~\cite{lin2018all,luo2022computational,mengu2022all}. Here our aim is to transform the speckle pattern from the MMF to a Gaussian beam with $M^2$ close to 1. The speckle pattern contains phase singularities~\cite{senthilkumaran2005vortex, de2016spatial}, where light intensity vanishes. If the original beam has no phase singularity, one phase-mask is sufficient for arbitrary intensity conversion~\cite{jesacher2008near,scholes2020improving}. But with phase singularities, one additional phase-mask is needed to obtain a singularity-free intensity profile~\cite{barre2022holographic}. A third mask is used for transformation to the final phase distribution.  

    Numerically we try three phase-masks that are equally spaced along $z$-axis. The axial distance $d_p$ between them is chosen such that free-space propagation from one mask to the next will result in sufficient mixing of phase-modulated light via diffraction~\cite{yildirim2024nonlinear}. We simulate light propagation and phase modulation in a forward model. After passing the third phase-mask, the optical beam propagates a distance $d_p$ to the target plane at $z=z_T$. We compute the final beam propagation factor $M^2$~\cite{siegman1990new} from the near-field and far-field intensity distributions, and use it as the cost function for optimizing phase modulations on three masks. The transverse size of each mask is set by the optical beam diameter covering 98~\% of optical power. To suppress phase modulation beyond the mask, we add a regularization term $\sigma$ to our cost function:
    \begin{equation}
    \sigma_o=M^2+\alpha \, \sigma.
    \label{eq:cost_output}
    \end{equation}
    $\sigma= \sqrt{ {\sum_{i=0}^{N_o}{ (\phi_i - \bar{\phi})^2 }} /{N_o} }$, where $\phi_i$ describes phase modulation at spatial position $i$ outside the beam diameter, and $N_o$ is the total number of spatial points. $\sigma$ characterizes the spatial variation of $\phi_i$ around its mean $\bar{\phi}$ outside the beam diameter. The weight of $\sigma$ in the cost function can be tuned by $\alpha$.
    For any output field $E(x,y,z_1)$ of the MMF, we iteratively search for the phase modulations on three masks $\{\phi_1(x,y),  \phi_2(x,y), \phi_3(x,y)\}$ at axial positions $z= z_1, z_2, z_3$ using error back-propagation and gradient descent~\cite{ruder2016overview, barre2022inverse}. More detailed information on our optimization scheme is provided in \ref{sec:appendix1}.

    Figures~\ref{fig:fig1}b--f show the evolution of the intensity profile before three phase-masks at  $z=z_1, z_2, z_3$ and at the target plane $z=z_T$. Figures~\ref{fig:fig1}f--i are the corresponding phase profiles. The phase holograms on three masks, $\phi_1(x,y), \, \phi_2(x,y), \, \phi_3(x,y)$, are plotted in Figs.~\ref{fig:fig1}j--l. After passing the first and second phase-masks, the intensity profile of the optical beam incident to the third mask is already shaped to the target Gaussian beam, as seen in Fig.~\ref{fig:fig1}d. However, the phase profile in Fig.~1h still differs from the flat phase of a Gaussian beam. Therefore, a third phase-mask is applied to flatten the phase profile. At the target plane $z=z_T$, the phase is flat with a slight tilt (Fig.~\ref{fig:fig1}i), while the intensity profile is Gaussian (Fig.~\ref{fig:fig1}e). Consequently, the far-field intensity in Fig.~\ref{fig:fig1}m displays a smooth profile, and $M^2 = 1.17$. We find that adding the fourth optimized phase-mask will make $M^2$ even closer to 1, if desired.

    Our method works for arbitrary beam profiles emerging from an MMF, i.e., for any number of modes and arbitrary modal compositions. So far we have ignored polarization mixing in the fiber, and assumed that the output field is linearly polarized.
    Since it is straightforward to extend our method to the MMFs with polarization mixing, we briefly describe the scheme below without showing an numerical result. Consider a MMF with strong polarization mixing, the output field has a spatially varying polarization state. A polarizing beamsplitter (PBS) is placed at the fiber output end to separate the linear-polarization components along horizontal ($H$) and vertical ($V$) axes. They are separately transformed, each by three phase-masks, to the Gaussian beam of same spatial profile. Then the horizontally and vertically polarized Gaussian beams are recombined by another PBS. By adjusting the relative (global) phase between the two beams with the third phase-masks in their paths, we can control the polarization state of the final Gaussian beam.
    
 \section{Spectral correlation width}

   \begin{figure*}[!h]
		\centering
  \includegraphics[width=1\textwidth]{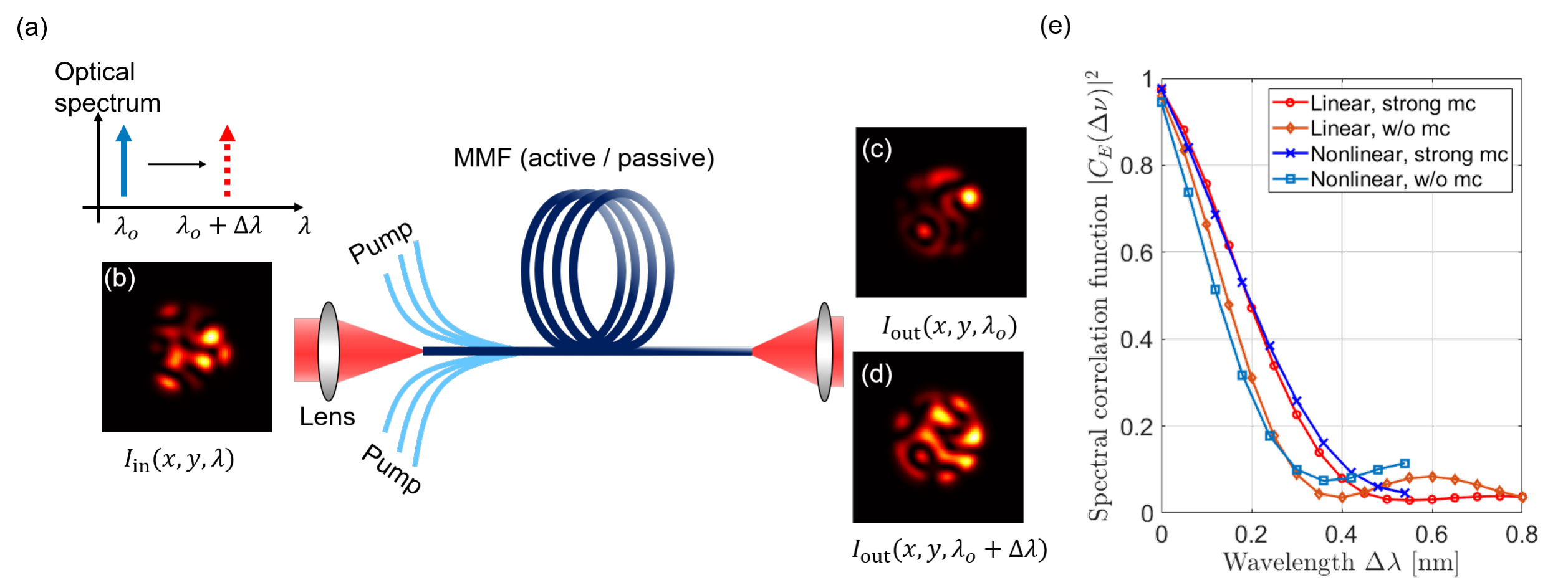}\\
		\caption{Spectral correlation width of a nonlinear MMF amplifier. We simulate 10~W signal at 1064 nm and 500 W pump at 976 nm coupled into a Yb:doped MMF amplifier. The fiber is 2 meter long, and has core diameter of 42~\textmu m and NA of 0.08, supporting 25 guided modes per polarization. At fiber output, signal is amplified to 444~W and pump depleted to 4.4~W . (a) Schematic showing the spectral correlation of a MMF with all modes excited equally. The input field pattern of signal is fixed and the wavelength is tuned from $\lambda_0$ to $\lambda_0 + \Delta \lambda$, resulting in decorrelation of output field pattern, characterized by  Pierson correlation $C_E(\Delta \lambda)$.  (b) Intensity distribution of the input signal, which remains constant with wavelength tuning. (c,d) Output intensity distribution at $\lambda_o = 1064.000$~nm and $\lambda_o + \Delta\lambda = 1064.045$~nm for the passive MMF with negligible mode coupling~(mc). (e) Spectral correlation function $|C_E(\Delta \lambda)|^2$ decays with wavelength detuning $\Delta \lambda$. Its width is nearly identical in MMFs with and without gain, reflecting negligible effects of gain saturation and pump depletion. Strong mode coupling increases the spectral correlation width in both passive and active fibers. }
		\label{fig:fig2}
	\end{figure*}
 
	In the previous section, we consider single-frequency input to a MMF. The output field has spatially varying phase, but the relative phase between any two positions is time invariant. In this sense, the output light is spatially coherent, allowing it to be transformed to a Gaussian beam with $M^2 \approx 1$ by spatial phase modulation. In reality, any input light has a finite spectral bandwidth $\delta \lambda$. If different spectral components create distinct field patterns at the output of a MMF, it would be impossible to use the same set of phase-masks to convert all of them to a Gaussian beam~\cite{van2011frequency}
 . However, if the input bandwidth is sufficiently narrow, the output field patterns for individual spectral components are nearly identical, then our method of beam transformation as described in the last section still works~\cite{lee2023efficient}. Now the question is how narrow the input bandwidth should be.
	
	The requirement for the input bandwidth of a MMF depends on how fast the output field pattern decorrelates with frequency detuning when the input wavefront is fixed. The latter is quantified by the spectral correlation function defined below~\cite{redding2013all}. Consider the fiber modes are excited by a fixed input wavefront, and the wavelength is tuned from $\lambda_0$ to $\lambda_0 + \Delta \lambda$ (Fig.~2a). The MMF output field changes from ${\bf E}(x,y,z; \lambda_0)$ to ${\bf E}(x,y,z; \lambda_0 + \Delta \lambda)$, here ${\bf E}$ denotes the vector field including both polarizations. The spectral correlation function is defined as	
	\begin{equation}
	C_E(\Delta \lambda) \equiv \frac{\langle {\bf E}^*(x,y,z; \lambda_0) \cdot {\bf E}(x,y,z; \lambda_0 + \Delta \lambda) \rangle}{\langle \left| {\bf E}(x,y,z; \lambda_0) \right|^2 \rangle} \, ,
	\end{equation}
	where $\langle ... \rangle$ represents averaging over spatial position $(x,y,z)$. 
The spectral correlation width $\Delta \lambda_c$ is given by the  full-width-at-half-maximum (FWHM) of $|C_E(\Delta \lambda)|^2$. If its bandwidth $\delta \lambda \ll \Delta \lambda_c$, the input light can be treated as monochromatic. For a passive MMF without mode coupling, $\Delta \lambda_c$ depends on the excitation condition. If all modes are equally excited at the fiber input, $\Delta \lambda_c$ scales as $1/L/\rm{NA}^2$, where $L$ is the fiber length, and $\rm{NA}$ is the numerical aperture (NA) of MMF. Strong mode coupling in a MMF will change the length scaling from $1/L$ to $1/\sqrt{L}$ \cite{ho2011statistics}.

    In an active MMF, gain saturation will induce nonlinear mode coupling. Thus the spectral correlation width could, in principle, deviate from that of a passive fiber. We numerically calculate $\Delta \lambda_c$ of a co-pumped Yb:doped MMF amplifier~\cite{paschotta1997ytterbium}. We consider a 2-meter-long step-index fiber with 42~\textmu m core diameter and 0.1 NA~\cite{wisal2024theoryTMI}. The signal at 1064 nm is amplified while propagating in the fiber, simultaneously the pump at 976 nm decays. The Yb:doped fiber parameters~\cite{smith2011mode} are given in table~C.1 of Appendix~C.  We solve the coupled equations for signal and pump power to account for gain saturation and pump depletion. The saturated gain coefficient at the end of the fiber is roughly 50 times smaller than the linear gain coefficient, confirming strong gain saturation in the MMF. To simulate strong mode coupling, signal fields in all 25 guided modes are multiplied by unitary random matrices at the beginning, in the middle and at the end of the fiber. For comparison, we also compute $\Delta \lambda_c$ of the same MMF without gain. 
    
    Figure 2e shows that the spectral correlation width in a high-power MMF amplifier with strong gain saturation and pump depletion is quite close to the value in the same fiber without any gain, when the linear mode coupling is negligible or strong. This result can be understood as follows. If two frequencies are within the spectral correlation width for the passive fiber, the modal propagation constants at these two frequencies are close enough to produce a highly correlated intensity distribution throughout the fiber. As such, the spatial variations in saturated gain, which depend on the local intensity, are also highly correlated. Therefore, the amplified fields at these frequencies have similar spatial distribution, as long as their frequency difference is within the spectral correlation width for the passive fiber. Note that we have assumed the emission and absorption cross-sections of the gain medium are roughly constant over the spectral correlation width $\Delta \lambda_c$, which is consistent for realistic MMF amplifiers. 
     \begin{figure*}[t]
		\centering
  \includegraphics[width=0.98\textwidth]{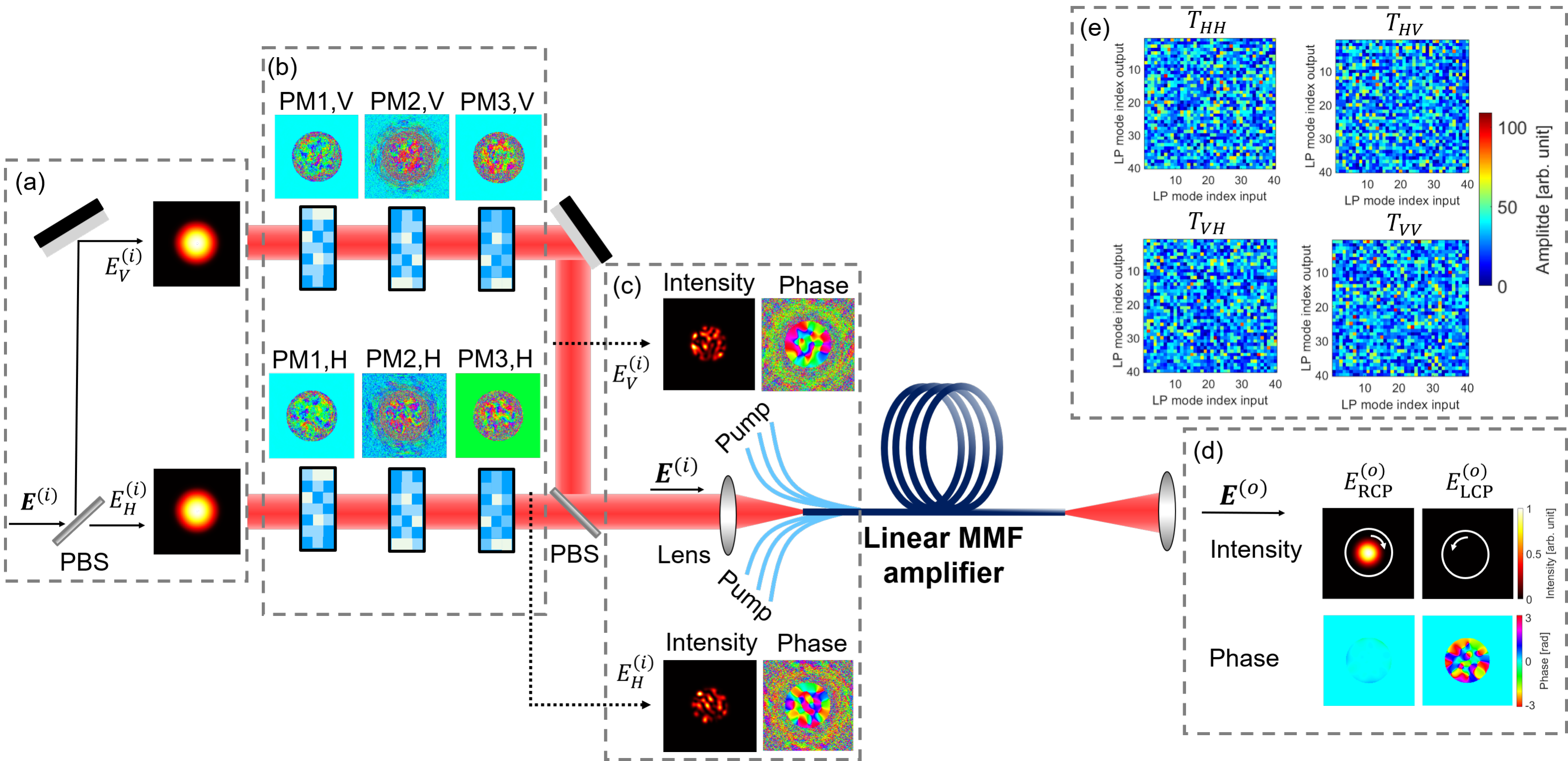}\\
		\caption{Input wavefront shaping of a linear MMF amplifier. The step-index fiber with 40 modes per polarization has linear mode-dependent gain, strong mode coupling and polarization mixing. (a) A Gaussian beam at 1064 nm is split by a polarization beam splitter (PBS) to horizontal (H) and vertical (V) polarizations. The power ratio of V to H is $p$. (b)  Each beam passes through three phase-masks (PM) of modulation patterns obtained by gradient descent optimization to have an output beam in the LP$_{\rm 01}$ mode. The global phase of the third PM of H relative to that of V is further tuned to set the output polarization state to right circular polarization (RCP).  (c) Shaped intensity and phase distributions for $H$ and $V$ polarizations. The two beams are recombined by another PBS before entering the MMF amplifier. (d) output field intensity and phase distributions of RCP and left circular polarization (LCP), confirming the output is a right-circularly polarized beam in the LP$_{01}$ mode with $M^2=1.08$. (e) Four sub-matrices $T_{HH}$, $T_{HV}$, $T_{VH}$, and $T_{VV}$ of simulate transmission matrix with strong random mode coupling, polarization mixing. } 
		\label{fig:fig3}
	\end{figure*}

 \section{Input wavefront shaping} 
 \label{sec:3}

 \subsection{Linear MMF amplifier}
     For a MMF amplifier, as long as the input bandwidth is much narrower than the spectral correlation width, the output beam remains spatially coherent and can be converted to a Gaussian beam with $M^2 \approx 1$ with a series of phase-masks, using the same method for passive MMFs in the previous section. To avoid high-power handling of amplified light at the fiber output, the phase-masks may be moved to the input end of a MMF amplifier to shape the spatial wavefront of a low-power seed. 
	
	We first consider a linear MMF amplifier with negligible gain saturation. In this case, we can rely on the field transmission matrix $T$ that maps spatial field distribution at the fiber input facet ${\bf E}^{(i)}(x,y)$ to that at the output facet ${\bf E}^{(o)}(x,y)$ \cite{florentin2019shaping}. 
	\begin{equation}
	\left[ {\begin{array}{cc}
		E_H^{(o)}(x,y)  \\
		E_V^{(o)}(x,y)
		\end{array} } \right]
	= \left[ {\begin{array}{cc}
		T_{HH} & T_{VH} \\
		T_{HV} & T_{VV} \\
		\end{array} } \right] 
	\cdot 	
	\left[ {\begin{array}{cc}
		E_H^{(i)}(x,y)  \\
		E_V^{(i)}(x,y)
		\end{array} } \right]
    \label{eq:T}
	\end{equation}
	$E_H^{(i)}(x,y)$ and 
	$E_V^{(i)}(x,y)$ denote the input field components of horizontal (H) and vertical (V) polarization respectively. $E_H^{(o)}(x,y)$ and $E_V^{(o)}(x,y)$ are output fields with H and V polarizations. For a MMF with $N$ modes, the $2N \times 2N$ transmission matrix $T$ consists of four $N \times N$ sub-matrices. For instance, $T_{HH}$ maps input H polarization to output H, and $T_{HV}$ maps input V to output H. Off-diagonal matrix elements of $T_{HH}$ and $T_{VV}$ represent mode coupling between spatial modes with same polarizations, whereas $T_{VH}$ and $T_{HV}$ describe polarization mixing in the MMF. $E_H^{(o)}(x,y) = T_{HH} \, E_H^{(i)}(x,y)+T_{ HV} \, E_V^{(i)}(x,y)$, and $E_V^{(o)}(x,y) = T_{VH} \, E_H^{(o)}(x,y)+T_{VV} \, E_V^{(i)}(x,y)$.  

    \begin{figure*}[h]
	\centering
\includegraphics[width=0.98\textwidth]{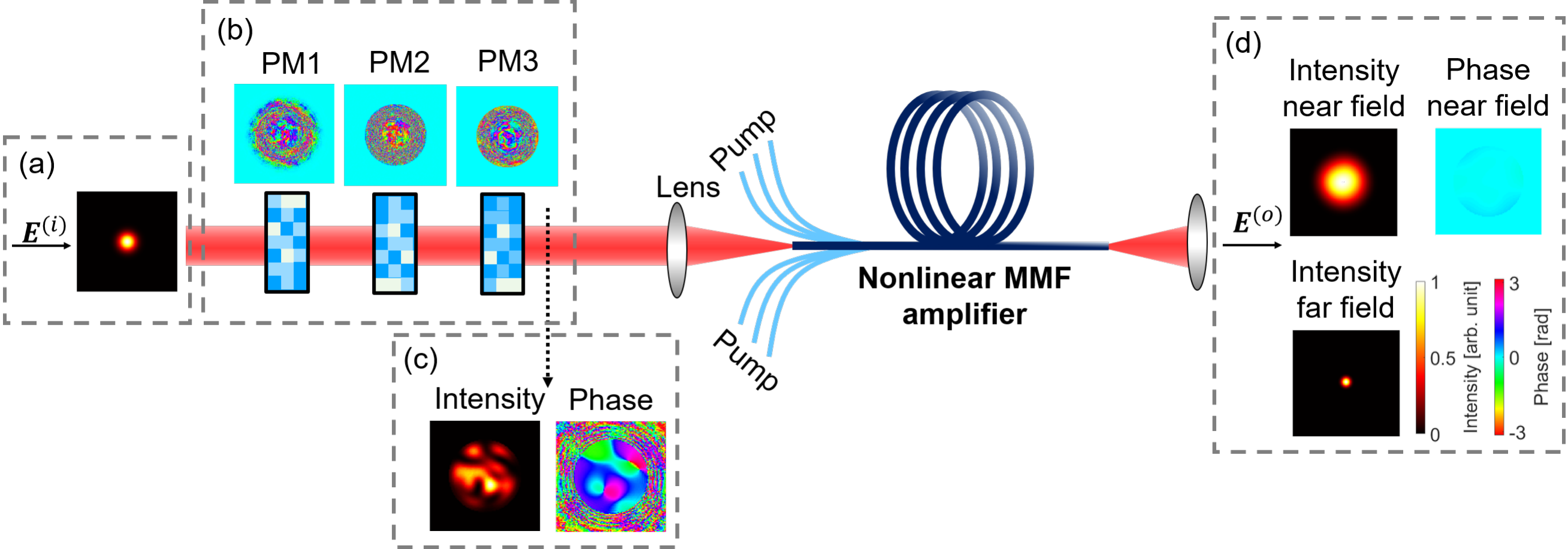}
		\caption{Input wavefront shaping of a nonlinear MMF amplifier. The step-index fiber of 42~\textmu m core and 0.06 NA supports 15 modes per polarization at signal wavelength 1064~nm. The input signal power is 10~W and pump power 500~W at 976~nm wavelength. The output signal power is 444.30~W and pump power 3.27~W. The saturated gain coefficient at the fiber output end is roughly 50 times smaller than the linear gain coefficient, which is nearly mode independent. Polarization mixing in the fiber is neglected.  
        A linearly-polarized Gaussian beam (a) passes through three phase-masks (b) before entering the MMF amplifier. The phase-masks in (b) are designed to have the amplifier output in LP$_{\rm 01}$ mode. (c) Spatial intensity and phase distributions of input signal to the MMF amplifier. (d) Output field intensity and phase distribution resembles the LP$_{01}$ mode, and far-field intensity profile is smooth. $M^2$ = 1.21 is close to 1. 
            }
		\label{fig:fig5}
	\end{figure*}
 
For a linear MMF amplifier with strong mode coupling and polarization mixing~\cite{ho2012exact,ho2011statistics, xiong2018complete, chiarawongse2018statistical}, the transmission matrix can be expressed as $T=U \, \Lambda \, V$, where $U$ and $V$ are $2N \times 2N$ unitary random matrices, and $\Lambda$ is a diagonal matrix of element $e^{g_i}$, $i = 1, 2, ... 2N$. The probability density of $g_i$ satisfies a semi-circle distribution for large $N$~\cite{ho2012exact}. We generate a $\Lambda$ with diagonal elements in the range $g_{\rm min}=10~dB$ and $g_{\rm max}=30~dB$, then multiply it with $U$ and $V$ to get $T$. The four $N \times N$ blocks of $2N \times 2N$ transmission matrix $T$ correspond to $T_{HH}$, $T_{HV}$, $T_{VH}$, and $T_{VV}$, as shown in Fig.~\ref{fig:fig3}e.

Once the transmission matrix of a MMF is computed, the matrix inverse $T^{-1}$ will provide the input field ${\bf E}^{(i)}(x,y)$ that is required to create a target output ${\bf E}^{(o)}(x,y)$. As an example, the target output beam is in the fundamental mode LP$_{01}$ with right circular polarization ${\bf E}_{\rm{RCP}}(x,y)$. We separate its field components of H and V polarizations, $\left[ {E}_H^{(o)}(x,y), {E}_V^{(o)}(x,y) \right]$.  They are multiplied by $T^{-1}$ to give the target input fields $\left[ {E}_H^{(i)}(x,y), {E}_V^{(i)}(x,y) \right]$, which are shown in Fig.~3c. 

We know from Sec.~2 that any field distribution can be created losslessly with three phase-masks. Here, we apply the same method to design three phase-masks for each polarization at the fiber input end, in order to create the target input fields ${E}_H^{(i)}(x,y)$ and ${E}_V^{(i)}(x,y)$. 
The cost function for our gradient descent algorithm is changed from $M^2$ to the deviation of shaped field distribution $\tilde{E}_{H,V}(x,y)$ and the target input ${E}_{H,V}^{(i)}(x,y)$: 
    $\bigl| {\tilde{E}_{H,V}(x,y) - E}_{H,V}^{(i)}(x,y) \bigr|^2$.
    We also include the term $\alpha \, \sigma$  as in Eq.~(\ref{eq:cost_output}) to suppress phase modulation outside the optical beam on phase-masks in the cost function:
    
    \begin{equation}
        \sigma_i=\bigl| {\tilde{E}_{H,V}(x,y) - E}_{H,V}^{(i)}(x,y) \bigr|^2 + \alpha \, \sigma,
        \label{eq:cost_input}
    \end{equation}
    
    Note that the power of target input field ${E}_H^{(i)}(x,y)$ may differ from that of ${E}_V^{(i)}(x,y)$. If the power ratio of V to H is $p$, the polarization state of an input Gaussian beam, ${\bf E}^{(i)}(x,y)= {\bf e}_H  E_H{(i)} + {\bf e}_V \sqrt{p} E_V{(i)}$, is adjusted to have the same power ratio for $E_V{(i)}$ and $E_H{(i)}$, where ${\bf e}_{H,V}$ is the unit vector in horizontal (H) or vertical (V) axis. The beam is then split by a PBS, and then launched to separate phase-masks for parallel wavefront shaping of H and V polarizations.   
    We further tune the relative global phase between the last phase-masks for H and V to control the output polarization state (Fig.~\ref{fig:fig3}b). At the fiber output end, the field is in LP$_{01}$ mode with right circular polarization, as targeted (Fig.~\ref{fig:fig3}d). We compute the $M^2$ =1.08, in agreement with the $M^2$ value for LP$_{01}$ in the MMF.

\subsection{Nonlinear MMF amplifier}

 Finally we consider a nonlinear MMF amplifier with strong gain saturation and pump depletion. In the nonlinear regime, field transmission matrix for input-output mapping no longer holds. Time-reversal symmetry, which is used to show the possibility of obtaining the $M^2 \simeq 1$ output beam with input wavefront shaping, also breaks down due to light amplification in the fiber. However, in a MMF amplifier with linear gain, the input wavefront to create a target output beam can still be found by considering a complementary MMF with absorption, in analogy to the coherent perfect absorber~\cite{chong2010coherent, wan2011time}. For a nonlinear MMF with gain saturation and thermo-optical nonlinearity, we have proved theoretically and confirmed numerically that input wavefront shaping can create any target output beam profile as long as the amplifier operates below the instability threshold and reaches a steady state~\cite{chen2024exploiting}. However, pump depletion was ignored in our previous study. For a high-power fiber amplifier, taking into account pump depletion would be necessary, as will be done below.  

To account for pump depletion, we numerically launch the residual pump light from a MMF amplifier together with phase-conjugated signal field into a virtual MMF with same structural parameters to the MMF amplifier. The optical gain coefficient is replaced by an absorption coefficient of identical magnitude, and gain saturation is replaced by the same degree of absorption saturation. The signal attenuates as it propagates through the absorbing MMF, while the pump power grows. The transmitted pump together with the phase-conjugated signal are injected into the nonlinear MMF amplifier, in order to generate the target output beam. As an example, we consider the amplifier output in LP$_{01}$ mode. Since the MMF amplifier has random mode coupling and mode-dependent gain, the complementary MMF absorber must have identical linear mode coupling and mode-dependent absorption. Nonlinear mode coupling, induced by saturated gain in the MMF amplifier, is reversed in the MMF absorber with equal amount of saturated absorption. To reduce the computation time, the number of guided modes is reduced to 15 by lowering the fiber NA. For simplicity,  polarization mixing in the MMF is neglected.

We find the input wavefront to the nonlinear MMF amplifier numerically by launching the amplified signal in $\rm LP_{01}$ mode together with the residual pump into the distal end of an equivalent MMF absorber. The transmitted signal at the fiber proximal end is phase conjugated and defines the input field profile to the MMF amplifier. It is then generated by three phase-masks placed at the amplifier input end. Figure~\ref{fig:fig5}b shows phase modulation patterns on three masks, obtained with the same optimization procedure and cost function [Eq.~(\ref{eq:cost_input})] as in the previous subsection. For confirmation, we simulate an input Gaussian beam passes through the designed phase-masks (Fig.~\ref{fig:fig5}c) and then couples into the nonlinear MMF amplifier. Fig.~\ref{fig:fig5}d shows the output beam in LP$_{01}$ mode, as targeted, with $M^2$ = 1.21. 

While the mapping from a nonlinear MMF amplifier to the complementary absorber confirms the existence of an input wavefront that creates an output beam with $M^2 \approx 1$, it is not practical to implement this mapping experimentally. By replacing the phase-masks with programmable spatial light modulators~(SLMs), the required phase modulations can be found by minimizing the deviation of measured output beam profile from the target one~\cite{florentin2017shaping, florentin2018space, florentin2019shaping}. Alternatively, the inverse mapping from output to input of a nonlinear MMF amplifier may be obtained by training an artificial neural network~\cite{genty2021machine,  xiong2020deep, teugin2020controlling,  pohle2023intelligent}. Experimentally, SLMs will create different input wavefronts and output beam profiles will be recorded. These data will be used for training a digital twin for the backward model that maps the amplifier output to input~\cite{wright2022deep, zhuge2023building}. After the training, the digital twin can predict the input field pattern for a target output. 
   
 \section{Discussion and conclusion}
	
Thermo-optical nonlinearity in the MMF amplifier, which was neglected in the last section, can be included by solving the heat diffusion equation~\cite{chen2024exploiting} coupled to the optical modal equations. We stress that our approach of utilizing highly multimode excitation in a MMF aims to mitigate nonlinear effects and increase instability thresholds. This would allow a high-power MMF amplifier to operate at steady state, which enables controlling the static output beam profile by wavefront shaping at amplifier input or output end. It avoids fast phase modulations that would be required above the instability threshold~\cite{montoya2017photonic}. Slow drift of fiber amplifiers can be compensated by replacing constant phase-masks with adaptive SLMs to adjust input wavefronts. Similar to the implementation of multiplane light conversion, multiple phase-masks may be implemented by multiple bounces from different regions of a single SLM, as the commercial SLM has more than one million pixels.

In this work, we have presented two wavefront shaping configurations. In the first one, phase-masks are placed at the MMF amplifier output end to directly shape the output beam. The advantage of this scheme is that additional phase-masks can be placed at the amplifier input end to optimize the multimode excitation in the fiber for maximal suppression of detrimental nonlinear effects such as SBS and TMI~\cite{wisal2024optimal}. Although phase-masks can operate at kW level~\cite{mohammadian2021versatile, Yoshiki2023kWSLM}, handling high-power beams is not easy and raises safety concerns. In the second scheme, phase-masks at the MMF amplifier input end are used to control the output beam profile. The advantage of this scheme is shaping low-power beams. However, if the mode coupling in the MMF is weak, creating an output beam in LP$_{01}$ mode means the input light will excite predominantly the LP$_{01}$ mode in the fiber. The quasi-single-mode excitation would weaken the suppression of SBS and TMI in the MMF amplifier. Focusing the output beam to a diffraction-limited spot outside but close to the fiber output facet will require excitation of most modes in the MMF, thus raising the power thresholds for SBS and TMI significantly, albeit the thresholds are not the highest possible.
    
While this work is focused on creating $M^2 \approx 1$ beams out of MMF amplifiers, our scheme of input wavefront shaping can be applied to generate other, more complex output patterns at high power, which is useful for material processing applications. The current method works for single-frequency or narrow-band MMF amplifiers. For a broadband MMF amplifier of spectral correlation width less than the signal bandwidth, spatio-spectral shaping~\cite{mounaix2020time, cruz2022synthesis, yessenov2022vector, cao2024spatiotemporal} will be needed for parallel wavefront shaping of different spectral channnels.

\subsection*{Acknowledgements}
	We thank David B. Phillips, Andr\'{e} D. Gomes, Tom\'{a}\u{s} \u{C}i\u{z}m\'{a}r, Peyman Ahmadi for stimulating discussions. This work is supported by the Air Force Office of Scientific Research (AFOSR) under Grant FA9550-20-1-0129 (H.C. and A.D.S.), the German Research Foundation (DFG) under Grant RO~7348/1-1~(S.R.) and the Austrian Science Fund (FWF) under Grant P 32146 (A.J.).
	
	
\subsection*{Competing interests}
	The authors declare no competing interests.
 
\appendix

\begin{figure*}[h]
	\centering
\includegraphics[width=0.9\textwidth]{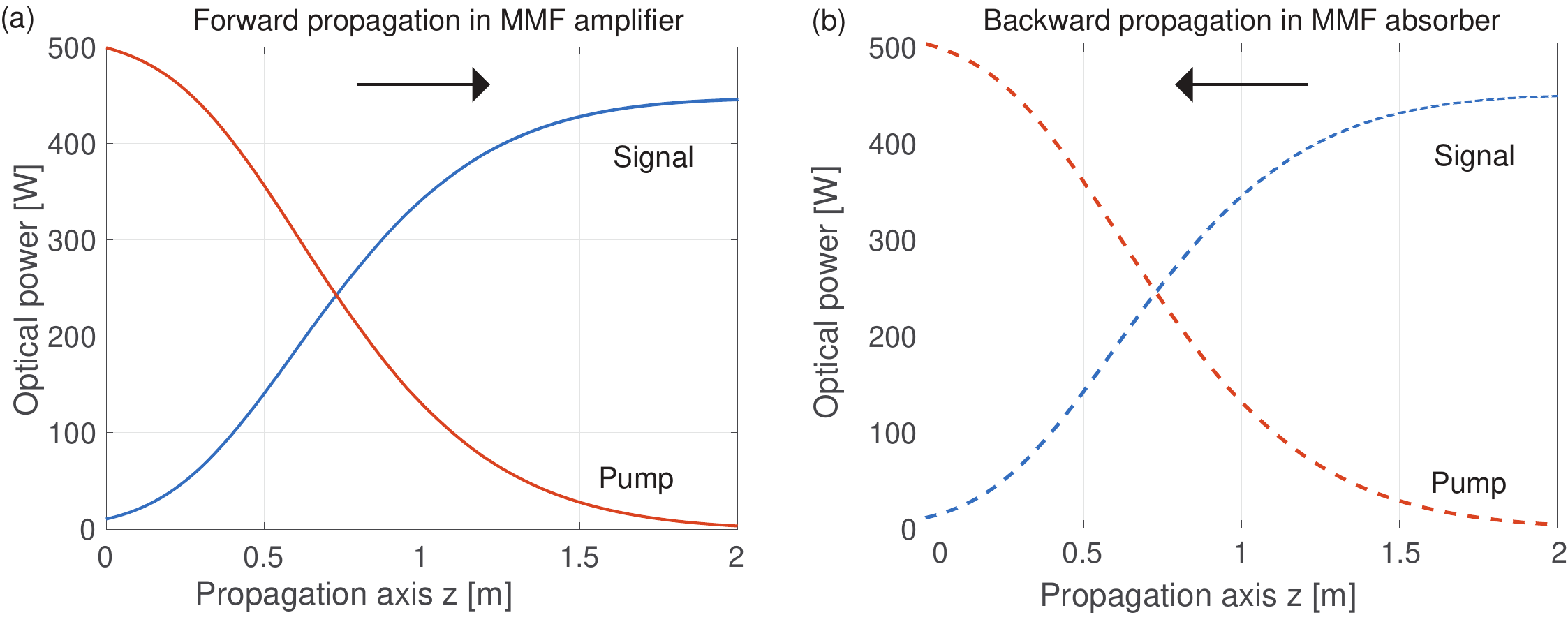}
		\caption{Nonlinear MMF amplifier and complementary absorber. (a) The pump and signal power evolution in a nonlinear MMF amplifier with parameters given in Table B1. The signal power (red solid) is amplified from 10 W to 444.3~W, and the pump power (blue solid) is depleted from 500 W to 3.27~W. (b) Output beam of the MMF amplifier is phase conjugated and launched to the distal end of the MMF absorber with the pump power of 3.27~W. In the backward propagation, changes of signal (red dashed) and pump (blue dashed) powers are exactly reversed to the MMF amplifier. The pump and signal powers at the distal end of the MMF absorber are equal to the input pump and signal powers of the amplifier. 
            }
		\label{fig:fig5}
	\end{figure*}
 
\section{Phase mask design}
\label{sec:appendix1}

Here we provide more detail on gradient-descent-based optimization of output wavefront shaping in Sec.~\ref{sec:sec2}. Three phase-masks are placed at the MMF output end. Optical beam profiles at these masks are computed on a grid of $100\times100$ pixels, with 10 pixels over the smallest feature size per dimension to ensure sufficient spatial resolution. Fields propagate in free-space from one phase-mask to the next, and optical diffraction leads to mixing of modulated fields. The distance between phase-masks is chosen such that light from one side of the beam will reach the opposite side~\cite{yildirim2024nonlinear}. To avoid spatial aliasing caused by diffracted light reaching the grid boundaries, we add $100$ pixels per dimension for zero padding that yields a total grid of $200\times200$ pixels. 

Given the incident field $E(x,y,z_1)$ on the first phase-mask, we iteratively search for the phase modulation patterns on all three masks, $\phi_i$ with $i = 1, 2, 3$, by backpropagating the cost function $\sigma_o$ in Eq.~(\ref{eq:cost_output}) through the masks. In each iteration $k$, we calculate the gradient $g_{k,\phi_i}$ for each phase-mask and update them via 
    \begin{equation}
        \phi_{k+1,i}=\chi\phi_{k,i}-\beta g_{k,\phi_i}.
        \label{eq:update}
    \end{equation}
We set the momentum $\chi=0.9$~\cite{ruder2016overview}. The learning rate $\beta$, which determines the step size during the search, is adjusted for the specific cost function.

\section{Simulation of nonlinear MMF amplifiers}
\label{sec:appendix2}

Here we provide more details on our numerical simulations of the nonlinear MMF amplifiers in sections 3 and 4.1. A detailed formulation of our method can be found in Ref.~\cite{wisal2024theoryTMI}.  We consider a co-pumped Yb-doped fiber amplifier with parameters given in the table. The signal beam propagating in the fiber core undergoes amplification. The growth of signal field amplitude in each fiber mode, $A_m$, is given by:

\begin{equation}
\frac{dA_m(z)}{dz} = \sum_{n=1}^N e^{i(\beta_n-\beta_m)z} g_{mn} (z) \, A_n (z) \, ,  
\label{Eq:signalamp}
\end{equation}
where the gain coefficient
\begin{equation}
g_{mn}(z) =  \frac{g_0}{2}\int d \vec{r}_\perp \frac{\psi^*_m (\vec{r}_\perp)\psi_n (\vec{r}_\perp)}{1 + I_0 (\vec{r}_\perp,z)/I_{\rm sat} (z)} \, .
\label{Eq:gaincoeff}
\end{equation}
Here, $\beta_m$ is the propagation constant of fiber mode $m$, $g_0$ is the linear (unsaturated) gain coefficient, $I_0 =| \sum_m A_m \psi_m|^2$ is the signal intensity at spatial position $(\vec{r}_\perp,z)$, $I_{sat}$ is the gain saturation intensity. $g_{mn}$ depends on the spatial overlap of transverse mode profiles $\psi_m$ and $\psi_n$, weighted by the intensity-dependent gain saturation factor, within the doped area of the fiber core. Notice that the cross gain terms ($g_{m\neq n}$) are non-zero due to the space-dependent gain saturation. As a result, the evolutions of different modal amplitudes are coupled, unlike the case where the gain saturation is absent and $g_{m\neq n} = 0$. The saturation intensity $I_{sat}$ and $g_0$ depend on the pump power $P_p (z)$, which decreases along the fiber:

\begin{equation}
\frac{dP_p(z)}{dz} = -g_p(z) P_p(z) \, , 
\label{Eq:pumpdecay}
\end{equation}
where $-g_p(z)$ is the pump absorption coefficient, and
\begin{equation}
g_p(z) = \frac{N_{\rm Yb}}{A_{cl}} \int d\vec{r}_\perp \left[ \sigma_p^a \, n_{\rm l}(\vec{r}_\perp,z) - \sigma_p^e \, n_{\rm u}(\vec{r}_\perp,z) \right] \, .
\end{equation}

The pump power decay rate varies along the fiber, as $g_p$ is proportional to the density of Yb atoms, $N_{\rm Yb}$, and depends on the fraction of atoms in upper ($n_u$) and lower ($n_l$) levels for amplification transition, as well as the absorption ($\sigma_p^a$) and emission cross-sections ($\sigma_p^e$) of Yb at the pump wavelength. The integral is over the doped area of the fiber (gain core), and $A_{cl}$ is the area of the pump core (inner cladding of the fiber). The upper ($n_u$) and lower ($n_l=1-n_u$) level population at steady state is obtained from the rate equations for the gain medium:

\begin{equation}
    n_{\rm u} = \frac{\frac{P_{\rm p}}{A_{\rm cl}\hbar\omega_{\rm p}}\sigma_{\rm p}^{\rm a}+\frac{I_0}{\hbar\omega_{\rm s}}\sigma_{\rm s}^{\rm a}}{\frac{P_{\rm p}}{A_{\rm cl}\hbar\omega_{\rm p}}(\sigma_{\rm p}^{\rm e}+\sigma_{\rm p}^{\rm a})+\frac{I_0}{\hbar\omega_{\rm s}}(\sigma_{\rm s}^{\rm e}+\sigma_{\rm s}^{\rm a})+\frac{1}{\tau}} \, .
    \label{Eq:nu}
\end{equation}

\begin{table}[t]
    \centering
    \begin{tabular}{|c|c|}
 \hline
 Parameter & Value \\ [0.5ex] 
 \hline\hline
 Core diameter [$\mu$m] & 42 \\ 
 \hline
 Cladding diameter [$\mu$m] & 570\\
 \hline
 Numerical aperture & 0.06, 0.08, 0.1\\
 \hline
 Signal wavelength [nm] & 1064\\
 \hline
 Pump wavelength [nm] & 976 \\
 \hline
 Pump absorption $\sigma^a_p$  [$\rm m^2$] & 2.47 $\times$ $10^{-24}$\\ 
 \hline
    Pump emission $\sigma^e_p$  [$\rm m^2$] & 2.44 $\times$ $10^{-24}$\\ 
 \hline
 Signal absorption $\sigma^a_s$  [$\rm m^2$] & 5.8 $\times$ $10^{-27}$\\ 
 \hline
 Signal emission $\sigma^e_s$  [$\rm m^2$] & 2.71 $\times$ $10^{-25}$\\ 
 \hline
 Dopant density $\rm N_{Yb}$ & 6.5 $\times$ $10^{25}$\\
 \hline

    \end{tabular}
    \caption{Table of MMF amplifier parameters}
    \label{fig:B1}
\end{table}

\noindent
Here, $\omega_{\rm s}$ and $\omega_{\rm p}$ are the signal and pump frequencies and $\sigma_{\rm s}^{\rm e}$ and $\sigma_{\rm s}^{\rm a}$ are the emission and absorption cross sections of the signal respectively. 
$\tau$ is the upper-level lifetime. The spatial dependence of $n_{\rm u}$ results from the variation of the pump power $P_{\rm p}$ and the signal intensity $I_0$ across the fiber. We solve Eq.~\ref{Eq:signalamp},  Eq.~\ref{Eq:pumpdecay} and Eq.~\ref{Eq:nu} together with the formula for $n_u$ using a finite-difference method such as Euler method.

We use this method to calculate the pump power and signal power across the nonlinear MMF amplifiers. Figure~\ref{fig:B1}a shows the pump and signal power evolution for an input pump power of 500~W and signal power of 10~W. The pump light gets absorbed while propagating along the 2-m-long fiber. It power decays to 3.27~W at the fiber output end. The signal power grows along the fiber, first exponentially with propagation distance, then linearly due to gain saturation, and eventually levels off as the pump is nearly depleted.

\section{Simulation of MMF absorber}
\label{sec:appendix2}

In this section, we provide more details on the backward propagation of signal and pump in the complementary MMF absorber. The parameters of the absorbing MMF are obtained simply by switching the sign of all emission and absorption cross-sections for the signal and the pump in Eq.~\ref{Eq:nu}. An alternative way, which we use, is to switch the sign of the dopant density $N_{\rm Yb}$ instead. The input signal power $P_{s1}$ and pump power $P_{p1}$ to the MMF absorber are equal to the output signal power and the residual pump power out of the MMF amplifier, respectively. 

Once we set the input signal power $P_{s0}$ to a nonlinear MMF amplifier and the target output signal power $P_{s1}$, we adjust the input pump power $P_{p0}$ in the simulation of forward propagation. We assume all modes are excited equally by the seed signal. The residual pump power at the amplifier output is used for the the input pump power $P_{p1}$ to the MMF absorber. We phase-conjugate the desired output beam profile of the MMF amplifier and launch it to the distal end of the MMF absorber. We simulate the backward propagation with initial signal power $P_{s1}$ and pump power $P_{p1}$. The transmitted signal beam at the proximal end of the MMF absorber is phase conjugated and defines the input wavefront to the MMF amplifier to generate the target output beam profile. Although such input wavefront creates the multimode excitation different from the equal mode excitation, the residual pump power out of the MMF amplifier barely changes. Figure~\ref{fig:B1}b shows the signal and pump power evolution during backward propagation in the MMF absorber. Compared to the MMF amplifier, the changes of signal and pump power are exactly reversed, leading to the original input pump and signal powers in the amplifier.

\bibliographystyle{elsarticle-num-names} 
\bibliography{cas-refs}





\end{document}